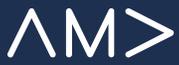

# 2023
# AMA WINTER ACADEMIC CONFERENCE

**Marketing During Times of Change**

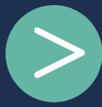 February 6 • Virtual
February 10-12 • Nashville, TN

## PROCEEDINGS

**Volume 34**

Co-Chairs:

**Nita Umashankar**
San Diego State University

**Monika Lisjak**
Arizona State University

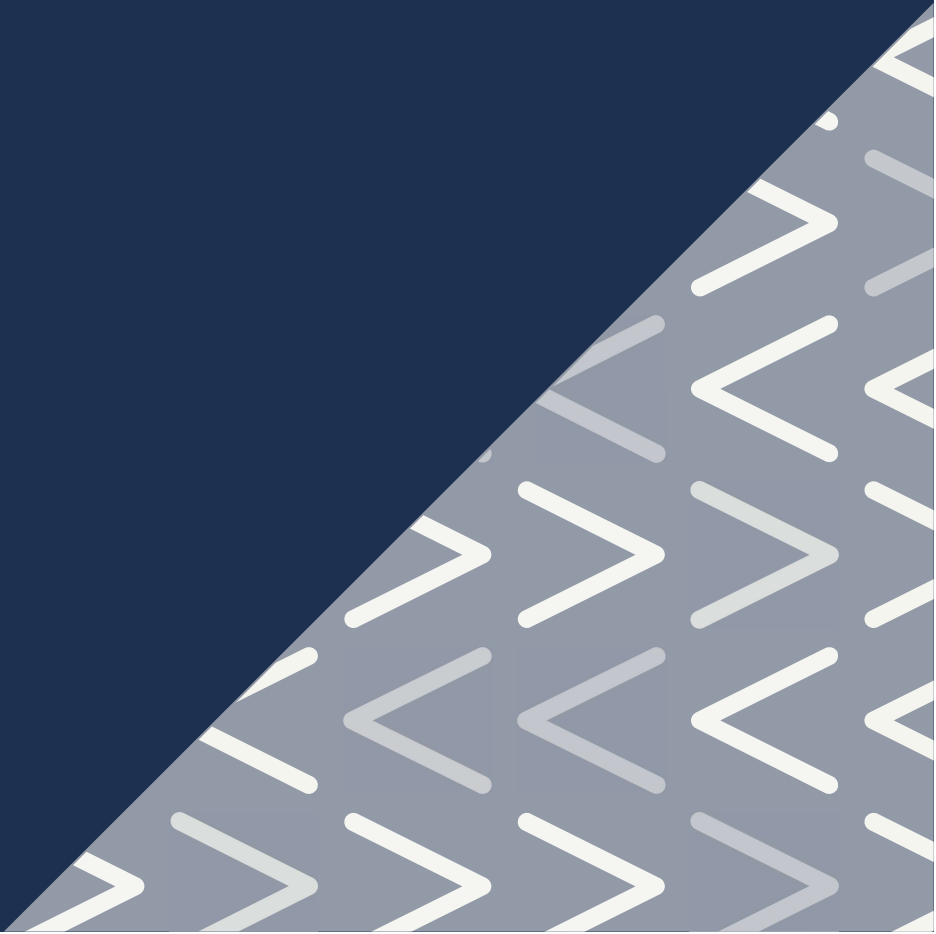

# 2023 AMA Winter Academic Conference

## Academic Conference

**Reconnecting and Reconceiving the Marketplace**
February 6 | Virtual
February 10-12 | Nashville, Tennessee

## PROCEEDINGS

**Volume 34**

Editors
Nita Umashankar, San Diego State University
Monika Lisjak, Arizona State University

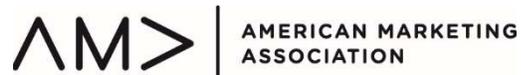

# "HOW CUSTOMERS' SATISFACTION CHANGE WITH THE USE OF AR SHOPPING APPLICATION:

## A CONCEPTUALL MODEL "

Fariba Sanaei, University of Central Florida

For further information, please contact Fariba Sanaei, PhD student, University of Central Florida

(fariba.sanaei@ucf.edu)

Description: This study provides a conceptual model of how online businesses using AR technology in their purchase stage can have effects on customer satisfaction and their purchase intention, considering the mediation roles of customer experience and immersion.

ABSTRACT

The paper proposes a conceptual model of how different perceived levels of experiential AR application features have effects on customer experience, and in turn their satisfaction and purchase behavior. In addition, it put forward the mediation role of immersion between perceived levels of experiential AR application features and customers' experience.

**INTRODUCTION**

The world economy experienced dramatic up and downs after Covid-19 pandemic (Ostrom et al., 2021). Before COVID-19, ecommerce was rapidly expanding. However, the epidemic drove even more US customers online. According to Digital Commerce 360 projections, the coronavirus contributed $102.08 billion to US ecommerce in 2020 and $116.45 billion in 2021. On the other hand, innovative technologies like AR have entered to marketing discipline and business owners found these technologies useful for attracting consumers. As a result, nowadays more businesses are turning into AR technology to increase their sales. adopting such strategies for online shopping enhances the sense of real environment in digital platforms and it can increase consumers engagements in long-term.

The goal of this study is to understand how adopting AR technology by business owners could be effective in terms of their consumer's purchase intention and their levels of satisfaction. By doing a comparative study in these field, beneficial implications can be received by business owners and academics. AR technology has entered in different areas from fashion industry to beauty cosmetics. Worldwide brands like Ikea and Amazon utilizing AR services in their online selling strategies for their home furniture. It can be anticipated that by introducing Metaverse world, more and more businesses will turn to such innovative technologies, so more research is needed for better and in depth understanding of relation between adopting AR technology, consumers' satisfaction, and their purchase intention. This study is going to answer: How is the consumers' experience of using AR applications affected, given different perceived level of experiential AR application features? How different perceived experience can result in different satisfaction levels and purchase intention?

**CONCEPTUAL FRAMEWORK**

Consumer experience is in the center of our framework. According to Lemon and Verhoef (2016), customer experience is defined by their responses during their shopping journey and these responses consist of multiple constructs like customers' cognitive, behavioral, emotional. in our research the main focus is on behavioral construct of consumers' experience. The following concepts are our main framework aspects for adopting AR technology buy business owners in their marketing strategies. 1) immersion defined by how an AR technology can reflect the realism with the use of different sensory feelings. 2)consumer experience as we defined before, and our main focus is on their behavioral responses. 3) experiential context defined as one of the constructs which identifies the positive points of adopting AR technologies for enhancing consumers' experience. 4) effectiveness of experiential AR application features including informative, personalizing, and interactivity features along the customers' experience and purchase stage of customers' journey.

***Stimulus-Organism-Response Paradigm***

By using the stimulus-organism-response S-O-R paradigm, retailers are able to understand the needed stimuli and their effects on consumers' behavior. This paradigm states that stimuli activate the cognitive processing of individuals and result in consumer responses like accepting or rejecting them (Mehrabian and Russell, 1974). One of the cases where this paradigm can be adopted is with AR mobile applications investigating app features and consumer behavior. There are studies that examined different important mobile AR app characteristics that have consequences on users' reactions like purchase intention (Daassi and Debbabi, 2021). Our study can add value to the literature of mobile AR application concepts by understanding how AR application features are experienced and perceived by consumers. The features that are going to be studied in this research include information, personalized and interactivity features.

Different consumers' experiences have been studied in previous research, like interactivity by Poushneh and Vasquez-Parraga in 2017, and informativeness by Rese et al. in 2017, etc. We focus on Javornik's perspective (2016) and consider the perceived levels of the informative, personalization, and interactivity features. For better understanding, the informative aspect of these applications enhances the process of decision-making for consumers and in our research refers to the level of the product information provided by AR app which is perceived by consumers. This information can be in the form of text, images, videos, etc. (Pantano et al., 2017).

Personalization features refer to perceived level of the information provided based on individuals' desires and needs according to their preferences. The presence of this feature utilizes customers with saving personal time, search costs, and money. By interactivity features, we mean different contents that people can have access to, and when this interactivity is high, the quality of AR is better, resulting in more positive responses from consumers. In this research we focus on perceived level of this feature by customers using the AR app, and this feature can be reflected into the technologically effective delivery process (Fiore et al., 2005). So, we propose the following hypotheses:

P1: The perceived level of informative feature of an experiential AR application positively influences customers' experience.

P2: The perceived level of personalizing feature of an experiential AR application positively influences customers' experience.

P3: The perceived level of interactivity features of an experiential AR application positively influences customers' experience.

*Immersion*

Immersion is a construct defined as how much a reality is near to its simulation made by technology using five human senses (Slater and Wilbur, 1997). According to Wedel et al. (2020) previous scholars investigated that how stimulated environments are immerse and if they get more immerse how consumers' behavior is going to be. Our main focus in this research is going to be on how vision sensory can affect the experiential effectiveness of available AR application features on consumers' experience. So, we can propose that:

P4: the effectiveness of experiential present AR application features on consumers' experience is mediated by immersion.

*Information System Success Model*

Information system success (ISS) determines users' behavior to understand their experience like satisfaction with using a service or product (Delone and Mclean, 2016; Schaarschmidt and Höber, 2017). Augmented reality is a new concept in retail and adopting the ISS model can utilize researchers to investigate this technology and understand how consumers react and respond to it in a shopping context. The current study provides more information and specifics about the acceptance and use of augmented reality (AR) technology for home furniture in retail and online platforms. Previous research has shown that if the quality of this system is high, then customers are more willing to continue using this technology (Kim and Hwang, 2012; Van Pinxteren et al., 2019). Services adopting AR features are expected to enhance consumers' experience because they provide customized services for consumers, and as this customization increases, high levels of customer satisfaction will be reached (Kim and Hwang, 2012; Murali et al., 2016). Therefore, customer satisfaction can be enhanced in online contexts using AR technology (Quadri-Felitti & Fiore, 2013; Tom Dieck, Jung, Kim, & Moon, 2017). According to Srivastava and Kaul (2014), consumer experiences are the main factor for defining customer satisfaction. The amount of joy and pleasure experienced by consuming a product is reflected in the satisfaction term. It is defined as fulfilling desires and leads to a positive service or product experience (Chung et al., 2018). By doing this research, more value will be added to the consumer behavior literature. The level of satisfaction gained from using AR sheds light on the use of AR apps in the online furniture industry. Using this technology affects consumers' intention to purchase and customer returns (Keiningham et al., 2017; Poushneh and Vasquez-Parraga, 2017). So, we can propose the following hypothesis.

P5: consumers' experience using online shopping applications using AR technology positively influences consumers satisfaction during their purchase stage.

P6: consumers' experience using online shopping applications using AR technology positively influences purchase intentions during their purchase stage.

The proposed model is shown in figure 1.

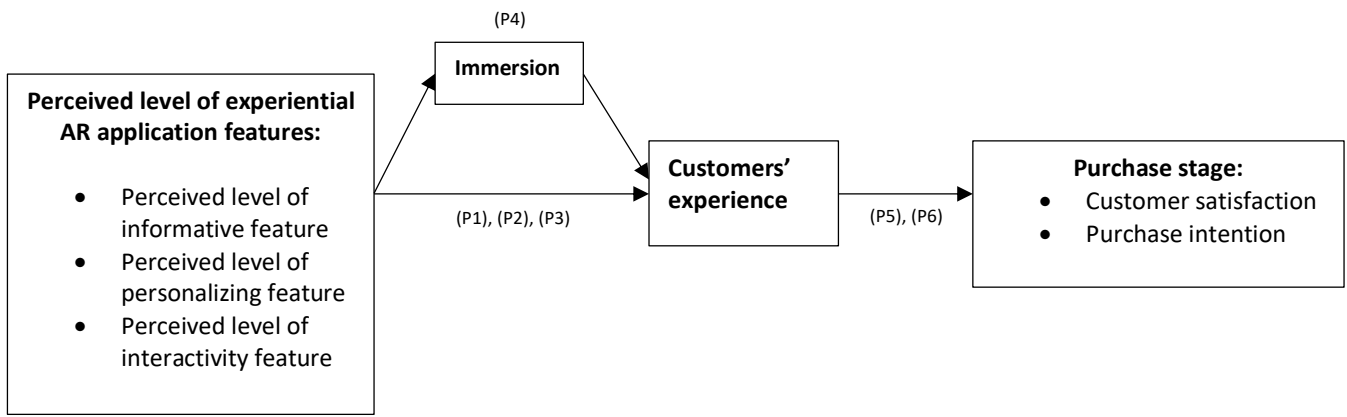

Figure 1. Conceptual Model

## DATA

The perceived level of experiential AR application features are subjective constructs and can be measured by effective delivery process techniques. Perceived informative features can be measured by Rese et al. (2017) methods. By adopting Kim and Baek (2018) methods, we can measure perceived levels of personalization of AR apps, and Pantano et al. (2017) proposed a method for measuring perceived interactivity feature and can be used and adopted for our research. The technical features and characteristics of AR applications like video inputs can be studied by subjective data to measure the immersion. Assessing the quality of these inputs can be measured by Akhtar and Falk (2017) method, and how they proposed the results can reflect the immersion of AR technology. In conclusion, for measuring the effectiveness of experiential AR application, immersion, consumers' experience, and consumers satisfaction, we can use primary data by asking consumers experiencing AR technology within questioners and qualitative interviews.

The experience scale developed by Brakus et al. (2009) can be adopted to assess behavioral aspect of consumers' experience. Self-report data from questionnaires are useful for capturing consumers' behavior. For the satisfaction, customer satisfaction measurement is a standard practice in marketing discipline, and it has been defined by the results of comparing perceived performance delivery with the customers' expectation. This confirmation or disconfirmation creates user satisfaction (Lemon and Verhoef, 2016). we are going to adopt Reichheld (2003) method of using Net Promoter Score (NPS) to gather our data for customer satisfaction.

For measuring purchase intention, at first, we can measure the number of sales in our study design which is done by our participants, furthermore for our final model, we can use secondary data from companies with AR technology for their online shopping applications, and asking them about the purchasing information of their consumers using their AR technology and information about the updates of their applications and specific changes in their sales according to these updates. This data set may be longitude for examining consumers' purchase intention.

**CONCLUSION**

In this article, we proposed that when online retailers adopting AR technology in their applications for customers' convenience, different levels of perceived experiential AR application features affect their experience. Informative, personalizing, and interactivity are three features that our study is going to investigate for their effects on users' experience. In addition, we proposed the mediating role of immersion between this perceived experiential and the customers' experience. We argued that the purchase stage behavior of a customer, like his or her satisfaction and purchase behavior, is influenced by his or her experience of using the retailers' AR technology.